# Noise-resilient quantum evolution steered by dynamical decoupling

Gang-Qin Liu[1#], Hoi Chun Po[2#], Jiangfeng Du[5], Ren-Bao Liu[2,3,4]*, Xin-Yu Pan[1*]

1.  *Beijing National Laboratory for Condensed Matter Physics and Institute of Physics, Chinese Academy of Sciences, Beijing 100190, China*
2.  *Department of Physics, The Chinese University of Hong Kong, Shatin, New Territories, Hong Kong, China*
3.  *Centre for Quantum Coherence, The Chinese University of Hong Kong, Shatin, New Territories, Hong Kong, China*
4.  *Institute of Theoretical Physics, The Chinese University of Hong Kong, Shatin, New Territories, Hong Kong, China*
5.  *Hefei National Laboratory for Physical Sciences at the Microscale and Department of Modern Physics, University of Science and Technology of China, Hefei, Anhui 230026, China*
# *These authors contribute equally*
* *Corresponding authors*

**Realistic quantum computing is subjected to noise. A most important frontier in research of quantum computing is to implement noise-resilient quantum control over qubits. Dynamical decoupling can protect coherence of qubits. Here we demonstrate non-trivial quantum evolution steered by dynamical decoupling control, which automatically suppresses the noise effect. We designed and implemented a self-protected controlled-NOT gate on the electron spin of a nitrogen-vacancy centre and a nearby carbon-13 nuclear spin in diamond at room temperature, by employing an engineered dynamical decoupling control on the electron spin. Final state fidelities of 0.91 and 0.88 were observed even with imperfect initial states. In the mean time, the qubit coherence time has been elongated by at least 30 folds. The design scheme does not require that the dynamical decoupling control commute with the qubit interaction and works for**

**general systems. This work marks a step toward realistic quantum computing.**

To combat with the noise effects in quantum computing, there are three main strategies, namely, quantum error correction [1-3], decoherence-free subspace[4,5], and dynamical decoupling (DD)[6-9]. DD, which is originated from magnetic resonance spectroscopy, can average out the noise by flipping the qubit back and forth. DD has the merits of requiring no extra qubits and potential compatibility with quantum gates. Recent experiments have demonstrated the protection of quantum coherence storage[10-15], or the NULL gate in terminology of quantum computing.

Integration of DD with quantum gates, however, is a nontrivial challenge since in general the quantum gates may not commute with the DD control and therefore can interfere with the DD. A straightforward approach is to insert the quantum gates in between the DD control sequences[16,17], which, however, significantly reduces the time windows for quantum gates. A clever solution is to design the DD sequences such that they commute with the qubit interaction, which can be realized either by encoding the qubits in decoherence-free subspaces[18] or by choosing a certain qubit interaction that commutes with the DD sequences[19]. It is also possible to apply control over one qubit while the other qubits are under DD control[20-22]. These methods, however, require special types of interactions or hybridized operations at very different timescales.

A more important question is whether DD, instead of locking the quantum states of qubits as previously demonstrated, can steer non-trivial and noise-resilient quantum evolutions of qubits with generic interactions (i.e., not limited to interactions commutable with the DD control)[23]. Here we demonstrate the feasibility of this strategy by steering the quantum evolution of a hybrid qubit system with engineered

DD control, which simultaneously realizes a non-trivial two-qubit gate and coherence protection. We realized a self-protected controlled-NOT gate on the electron spin of a nitrogen-vacancy centre and a nearby carbon-13 nuclear spin in diamond, by employing an engineered dynamical decoupling control applied only on the electron spin. The scheme is motivated by the recent study of central spin decoherence in nuclear spin baths [24-28], which reveals that due to the quantum nature of the qubit-bath coupling, the quantum evolution of nuclear spins is actively manipulated by flipping the central electron spin. Such quantum nature of qubit-bath coupling has been previously utilized to realize control of nuclear spins by flipping the electron spin [29-32]. Here we show that by carefully engineering the timing of the electron spin flipping, one can steer a noise-resilient quantum evolution of interacting qubits simply by DD control.

The method presented here does not require that the qubit interaction commute with the DD control and in principle can be applied to general quantum systems. Recent research on single nuclear spin sensing by central spin decoherence [31,33-37] already demonstrated the quantum nature of coupling between a nitrogen-vacancy centre spin and remote nuclear spins. The approach of quantum gates by DD may be applied to those remote nuclear spins, which is useful for scalable quantum computing.

**Design of quantum gates by DD.**

To demonstrate the concept of quantum steering by DD, we consider a negatively charged nitrogen-vacancy (NV) centre in a type IIa diamond (with nitrogen concentration <10 ppb) under an external magnetic field **B**. The NV centre electron spin is coupled through hyperfine interaction to $^{13}$C nuclear spins, which has a natural abundance of 1.1%[39] (Fig. 1a). Lifting the degeneracy between $m = +1$ and $m = -1$ NV centre spin states by **B**, we encode the first qubit in the centre spin states $|m = 0\rangle$ and $|m = -1\rangle$. To simplify notation, in the following we denote $|m = 0\rangle = |0\rangle$ and $|m = -1\rangle = |1\rangle$. Furthermore, since the hyperfine interaction strength decreases rapidly with the distance between the nuclear spin and the NV centre, a proximal $^{13}$C spin can be identified by its strong hyperfine splitting in the optically detected magnetic resonance (ODMR) spectra. We encode another qubit in the $^{13}$C nuclear spin-1/2 states $|\uparrow\rangle$ and $|\downarrow\rangle$ (Fig. 1b), similar to the electron-nuclear spin register studied in Ref. 31.

The undesired coupling of this two-qubit system to the other $^{13}$C spins in the bath leads to loss of quantum information and therefore reduces the fidelity of the quantum operations. In particular, due to the large difference between the gyromagnetic ratios of the two types of spins, the electron spin decoherence occurs in a timescale that is shorter than the typical operation timescale of the nuclear spin qubit, which results in difficulty in realizing high-fidelity two-qubit operations.

The quantum dynamics of the two-qubit system is a propagator in the curved SU(4) operator space[38]. In a free evolution the system propagator follows the natural landscape in the operator space, but the uncertainty in the system propagator increases with time due to the coupling with the environment (Fig. 1c). While a conventional

DD scheme like the Carr-Purcell-Meiboom-Gill (CPMG) sequence can efficiently refocus the otherwise non-coherent evolution of the system propagator, the sequence in general corresponds to an unspecified two-qubit propagator unless resonant values of the total evolution time and the number of pulses are chosen[21,36] or additional manipulation on the nuclear spin is employed[21]. Simultaneously achieving coherence protection with gate implementation, our scheme can be intuitively understood as a systematic approach to identify a path in the operator space comprising only free evolution and centre spin $\pi$-pulses such that the path is self-protected and guides the system propagator from the identity to a desired two-qubit gate (Fig. 1d).

Since the timescale of interest is much shorter than the longitudinal relaxation time of the centre spin, the centre spin magnetic number $m$ remains a good quantum number and the Hamiltonian of the two-qubit system can be expanded in the basis of the centre spin eigenstates $|0\rangle$ and $|1\rangle$, given by

$H = \sum_m |m\rangle\langle m| (E_m + \boldsymbol{\omega}_m \cdot \boldsymbol{I}) = \sum_m |m\rangle\langle m| h_m$, where $E_m$ is the eigen-energy of the centre spin state $|m\rangle$, $\mathbf{I}$ represents the nuclear spin and $\boldsymbol{\omega}_m$ is the local field for the nuclear spin conditioned on the electron spin state $|m\rangle$. In general, when the centre spin state is altered, the nuclear spin will evolve under a different local field and therefore the nuclear spin evolution is conditional on the state of the centre spin[31,32,35,39-41]. When the angle $\varphi$ between $\boldsymbol{\omega}_1$ and $\boldsymbol{\omega}_0$ is non-zero (Fig. 1b), which is expected in a general setting [see Supplementary Information for more discussion], they represent different axes on the nuclear spin Bloch sphere and hence can be utilized to generate universal qubit operations conditioned on the electron spin qubit.

To be specific, we suppose the system is prepared in an initial state $|\Psi\rangle = \sum_m |m\rangle |\psi_m\rangle$. When a sequence of $N$ $\pi$-pulses is applied to the centre spin, the nuclear spin state $|\psi_m\rangle$ evolves to $|\psi_m'\rangle = u_m\{t_\alpha\}|\psi_m\rangle$ with $u_0\{t_\alpha\} = e^{-ih_\sigma t_N}\dots e^{-ih_0 t_2}e^{-ih_1 t_1}e^{-ih_0 t_0}$, where $t_\alpha$ is the time between the $\alpha$-th and the $(\alpha+1)$-th pulses, $\sigma = 0$ for $N$ being even and $\sigma = 1$ for $N$ being odd, and $u_1\{t_\alpha\}$ is similarly defined. This implies that the system propagator can be represented by $U\{t_\alpha\} = \sum_m |m\rangle\langle m| \otimes u_m\{t_\alpha\}$. One can vary the timing parameters $\{t_\alpha\}$ and engineer the system evolution such that $U\{t_\alpha\} \approx G$ for some desired two-qubit gates with the generic form $G = \sum_m |m\rangle\langle m| O_m$, where $O_m$'s are nuclear spin operators. Important examples with this form include the controlled-NOT gate ($C_eNOT_n$), nuclear spin single-qubit gates, centre spin phase gates as well as the two-qubit NULL gate. In general, it is non-trivial to solve $u_m\{t_\alpha\} = O_m$ exactly due to the high non-linearity and large number of variables in the problem. Yet, one can recast the design protocol into a maximization problem through studying the average two-qubit gate fidelity

$$\bar{F}\{t_\alpha\} = \int d\Psi Tr\left(U\{t_\alpha\}|\Psi\rangle\langle\Psi|U^\dagger\{t_\alpha\}G|\Psi\rangle\langle\Psi|G^\dagger\right),$$ where $|\Psi\rangle$ is a general pure two-qubit state and the integration is over the normalized uniform measure of the state space[42]. Since $\bar{F}\{t_\alpha\} = 1$ if and only if $U\{t_\alpha\} = G$, the gate $G$ is simulated by the system propagator when the timing parameters $\{t_\alpha\}$ are chosen to maximize $\bar{F}\{t_\alpha\}$. Such gate design can be achieved by using only DD sequences, where the notion of

DD can be understood as a set of criteria on the timing parameters $\{t_\alpha\}$ between the centre spin $\pi$ pulses. While a conventional $N$-pulse DD sequence (say the CPMG sequence) is characterized by just one timing parameter, namely the pulse delay time, the general DD criteria can be derived by studying the expansion of the coherence function [Supplementary Information]. Instead of using all the timing parameters to optimize protection of the centre spin coherence[43], one can relax some of the timing parameters by reducing the decoherence suppression order. The design procedure is summarized as

$$
\begin{cases}
\text{maximization of } \left[ F\left(\{t_\alpha\}\right) \right] \text{ with respect to a } N\text{-pulse sequence,} \\
\text{1st order DD (echo condition): } t_0 - t_1 + t_2 + \cdots + (-1)^N t_N = 0, \\
\text{2nd order DD (symmetrization): } t_n = t_{N-n}, \\
\text{3rd order DD,} \\
\cdots
\end{cases}
$$

As a demonstration of principles, we consider the first order DD criterion for coherence protection together with the symmetric requirement on the pulse sequence[44,45]. Note that the first order DD criterion can be intuitively understood as the spin echo condition, which is independent of the bath spectrum. The symmetric timing condition realizes the second order DD [44,45]. The DD constraints are explicitly introduced by considering sequences of the form $\{t_\alpha\}_{DD} = \{t_0, t_1, t_{2,} \ldots t_2, t_1, t_0\}$ with the echo condition $\sum_\alpha (-1)^\alpha t_\alpha = 0$ [Supplementary Information]. With such, we design DD sequences that execute the desired two-qubit gates by maximizing $\bar{F}\{t_\alpha\}_{DD}$ with respect to the $N$ independent timing variables.

It is straightforward to generalize the design to higher order DD for better noise resilience.

We demonstrate the feasibility of the design by considering an experimentally identified target $^{13}$C spin coupled to an NV centre spin. Applying an external magnetic field **B** with a fixed strength of 100 G and a variable orientation, we adopted a mapping gate described in Ref. 31 to polarize and read out the nuclear spin. The experimental parameters were extracted from the ODMR spectra of the NV centre and the free precession signal of the nuclear spin, which were $\omega_0 = 0.256(2)$ MHz and $\omega_1 = 6.410(2)$ MHz. The polarization and readout fidelities were enhanced by carefully adjusting the orientation of **B** such that $\varphi$ is tuned to be 90°, which corresponds to the optimal conditions for the mapping gate [Supplementary Information].

Various DD gate sequences, listed in Table S1, were designed based upon the obtained experimental parameters, and their performances are summarized in Figs. 2a-d. Five different two-qubit gates, namely the $C_eNOT_n$ gate (defined up to an additional $\pi/2$ phase shift of the centre spin)[21], the nuclear spin Hammard ($H_n$), Pauli-X ($X_n$) and Pauli-Z ($Z_n$) gates, and the two-qubit NULL gate, were designed using sequences with 4 to 14 pulses. The gate fidelity $\bar{F}\{t_\alpha\}$[42], which was maximized in the optimization process, were found to be at least 0.98 for the gates we considered, and the gate operation time $T_G$ ranged from 1.4 to 5 μs. To incorporate the coupling between the two-qubit system and the environment, we simulated the environment by a small bath consisting of 6 $^{13}$C spins aside from the target $^{13}$C spin. The coupling to the spin bath causes centre spin decoherence in a timescale of $T_2^* \approx 1.5\ \mu s$ under free induction decay (FID), which is consistent with the experimental condition. To characterize the performance of the designed gates ( $G$ ), we considered a typical system initial state $|\Psi_0\rangle = \frac{1}{\sqrt{2}}\big(|0\downarrow\rangle + |1\downarrow\rangle\big)$ and we supposed the bath was in the thermal state denoted by $\rho_{\text{bath}}$. We then calculated the total propagator $U_T$ under the pulse sequences by exact diagonalization, assuming perfect $\pi$-pulses and inter-pulse timing. The state fidelity $\mathsf{F}$ of the resultant state, defined by $\mathsf{F} = \text{Tr}\big(\rho_i \rho_s\big)$, where $\rho_i = G\,|\Psi_0\rangle\langle\Psi_0|\,G^\dagger$ is the ideal system density matrix and $\rho_s = \text{Tr}_{\text{bath}}\Big[\,U_T\big(|\Psi_0\rangle\langle\Psi_0|\otimes\rho_{\text{bath}}\big)U_T^{\,\dagger}\,\Big]$ is the simulated system density matrix, was found to be at least 0.95 for the sequences we considered. We also characterized the coherence protection ability of the DD gate sequences by performing the cluster-correlation expansion (CCE) calculation[40] with a larger bath of 44 $^{13}$C spins. Due to entanglement with the target spin, the centre spin coherence right after applying the DD gate sequence can be 0. For a fair comparison, therefore, we probe the centre spin coherence $\mathsf{L}(t)$[41] after performing each gate sequence twice, such that ideally one would obtain $\mathsf{L}(2T_G) = 1$ for the gates we considered. The coherence function was found to be at least 0.92 even when the total evolution time exceeds the FID decohernce timescale. We note that while in principle the average two-qubit gate fidelity can be improved to almost unity by using a larger number of pulses, this does not automatically guarantee a higher resultant state fidelity when the spin bath is incorporated, and in the experimental setting this can also introduce additional pulse errors.

As the gates are realized by applying a DD sequence, they are self-protected, which offers an integrated solution to achieve both control and noise-tolerance in quantum information processing. We note that when the hyperfine interaction is moderately strong, the gate speed of the nuclear spin gates constructed this way can be significantly faster than the conventional control by radiofrequency pulses[29,30].

**Experimental implementation.**

We experimentally demonstrate our scheme by executing a designed 4-pulse $C_eNOT_n$ gate sequence (Fig. 3a). The system was first initialized into the $|0\downarrow\rangle$ and $|1\downarrow\rangle$ states with initial state fidelities, defined as $\mathrm{Tr}\left(\rho_i\rho_s\right)$ with $\rho_i$ being the ideal and $\rho_E$ being the measured reduced density matrix, of 0.92 and 0.91 respectively (Figs. 3b and 3e). The particular gate demonstrated can be readily understood by studying the state trajectory on the Bloch sphere. As illustrated in Figs. 3c and 3f, the nuclear qubit traces out paths that would respectively leave it unchanged or flipped when the centre spin is prepared in the $|0\rangle$ or the $|1\rangle$ states. State tomography revealed that the state $|0\downarrow\rangle$ was left unchanged as it should, with a final state fidelity of 0.91 (Fig. 3d). For the initial state $|1\downarrow\rangle$, the sequence flipped the nuclear spin and the resultant state was $|1\uparrow\rangle$ with a final state fidelity of 0.88 (Fig. 3g).

Having demonstrated the quantum gate aspect of the sequence, we now turn to illustrate the coherence protection aspect of our design by applying the $C_eNOT_n$ gate DD sequence to an initial superposition state $|\Psi_0\rangle = \frac{1}{\sqrt{2}}\left(|0\downarrow\rangle + |1\downarrow\rangle\right)$. Without coherence protection, the electron spin coherence decays in a short time of $T_2^* = 1.54(9)$ μs in FID, and the coherence was completely lost in ~2 μs (Fig. 4a). With the DD gate sequence, however, the centre spin coherence was well protected during the course of gate operation, even though it lasts for more than twice of $T_2^*$ (Fig. 4b). The coherence protection capacity of the DD gate sequence is further illustrated in Figs. 4c-f. The designed pulse sequence was repeated for $N=2, 4, 8, 12$ times on the initial state $|\Psi_0\rangle$, and the Ramsey interference signal was measured after the total sequence time of 8 μs, 16 μs, 32 μs, and 48 μs, correspondingly. The presence of strong coherent oscillation after an interval longer than 30 times of $T_2^*$ demonstrates the robust protection effect of the DD gate design.

This work demonstrates a general approach to quantum information processing in which the quantum evolution of the system is engineered to perform decoupling from a large spin bath and execute a designated gate on the two-qubit system, and thereby simultaneously realizes high fidelity two-qubit gates and coherence protection. The approach developed here is applicable to other systems under a general setting and therefore serves as a solution towards achieving scalable and fault-tolerant quantum computing.

# Methods summary

A home-built scanning confocal microscope combined with integrated microwave devices is employed to initialize, control and read out the electron spin state. The ODMR spectra of the NV centre was fitted by a multi-peak Gaussian form and $\omega_m$ can be extracted from the resonant frequencies. Selective Rabi oscillation was driven between different states using weak microwave pulses. The experimentally achievable polarization along $\omega_1$ using the scheme in Ref. 31 was enhanced by applying an additional "kick out" pulse to remove the unwanted component to the unused electron spin subspace.

State tomography was performed following a scheme in Ref. 46. The transitions between $\left|0\downarrow\right\rangle \leftrightarrow \left|1\downarrow\right\rangle$ and $\left|0\uparrow\right\rangle \leftrightarrow \left|1\uparrow\right\rangle$ were driven by weak MW pulses of controllable phase, and the population and phase of the corresponding density matrix elements were mapped out from the Rabi nutation signal. The other density matrix elements were obtained by transferring the corresponding population into either one of these two measurable transitions.

We adopted a numerical optimization protocol similar to the one adopted in Ref. 47 for the optimization of the average two-qubit fidelity $\overline{F}\{t_\alpha\}$ subjected to the DD constraints (echo condition and symmetric sequence). In each optimization step we update $t_\alpha \rightarrow t_\alpha + \varepsilon\partial_{t_\alpha}\overline{F}$ subjected to the DD constraints, where $\partial_{t_\alpha}\overline{F}$ can be directly calculated by evaluating $\partial_{t_\alpha}U$. The spin bath was simulated by considering the $^{14}$N nitrogen host nuclear spin together with $^{13}$C nuclear spins randomly placed in the diamond lattice with the natural abundance of 1.1 %. While the hyperfine coupling between the centre spin and the $^{14}$N was modeled by established parameters, the coupling between the centre spin and the rest of the bath spins, as well as the coupling between bath spins, was assumed to be dipolar[28,41]. The simulated bath was selected to reproduce the value of $T_2^*$ in FID experiment.

Full methods and related references are included in the Supplementary Information.

**Acknowledgements:** This work was supported by National Basic Research Program of China (973 Program project No. 2009CB929103), the National Natural Science Foundation of China Grants 10974251, National Natural Science Foundation of China Project 11028510, Hong Kong Research Grants Council - National Natural Science Foundation of China Joint Project N_CUHK403/11, The Chinese University of Hong Kong Focused Investments Scheme, and Hong Kong Research Grants Council - Collaborative Research Fund Project HKU10/CRF/08.

**Author Contributions**: R.B.L. proposed the project and conceived the idea. H.C.P. designed the scheme and carried out the theoretical study. X.Y.P. and G.Q.L. designed the experiment. G.Q.L. carried out the experimental study. H.C.P. and G.Q.L. wrote the paper. All authors analyzed the data and commented on the manuscript.

**Competing financial interests** The authors declare no competing financial interests.

**Correspondence** and requests for materials should be addressed to R.B.L. (rbliu@phy.cuhk.edu.hk) or X. Y. Pan (xypan@aphy.iphy.ac.cn).

**Supplementary Information** is linked to the paper.

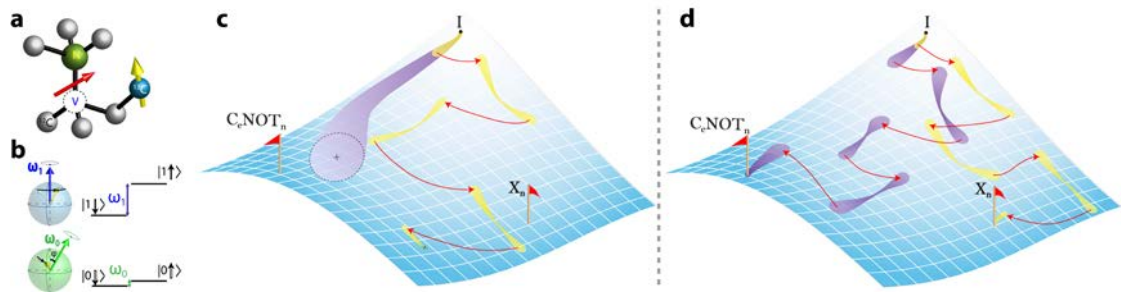

**Figure 1 | System and approach for steering quantum evolution by dynamical decoupling. a**, The electron-nuclear spin qubit system. The NV centre electron spin (electron qubit) is coupled to a bath of nuclear spins in a high purity diamond, and a proximal $^{13}$C nuclear spin (nuclear qubit) is identified by a stronger hyperfine coupling. **b**, The conditional local field experienced by the nuclear spin. Due to its coupling with the centre spin, both the quantization axis and the precession rate of the nuclear spin depend on the state of the centre spin. **c - d**, Quantum evolution in the operator space, starting as the identity operator I. Under a free evolution (yellow and purple path bundles), the propagator follows the natural "landscape" of the curved operator space. A stroboscopic control (the red arrows) moves the operator to a new position. **c**, Propagator evolution in the conventional settings. In an FID experiment (purple path), the coupling between the system and the bath causes non-ideal evolution of the system propagator, and therefore the path of the propagator spreads out in the operator space. In a conventional CPMG sequence (yellow path), $\pi$-pulses (red arrows) are applied at suitably timed intervals to decouple the system from the environment, and can therefore refocus the uncertainty in the system propagator. Yet, in both cases the quantum dynamics of the system is unconstrained and in general the resultant propagator is not the target one. **d**, Steering quantum evolution by DD. In

contrast to the conventional schemes, one can relax the timing constraint in applying the DD $\pi$-pulses and thereby steer the system propagator to the desired points in the operator space. For instance, the $C_eNOT_n$ and the $X_n$ gates can be respectively realized by the purple and the yellow paths by tuning the timing parameters in the 4-pulse DD sequence.

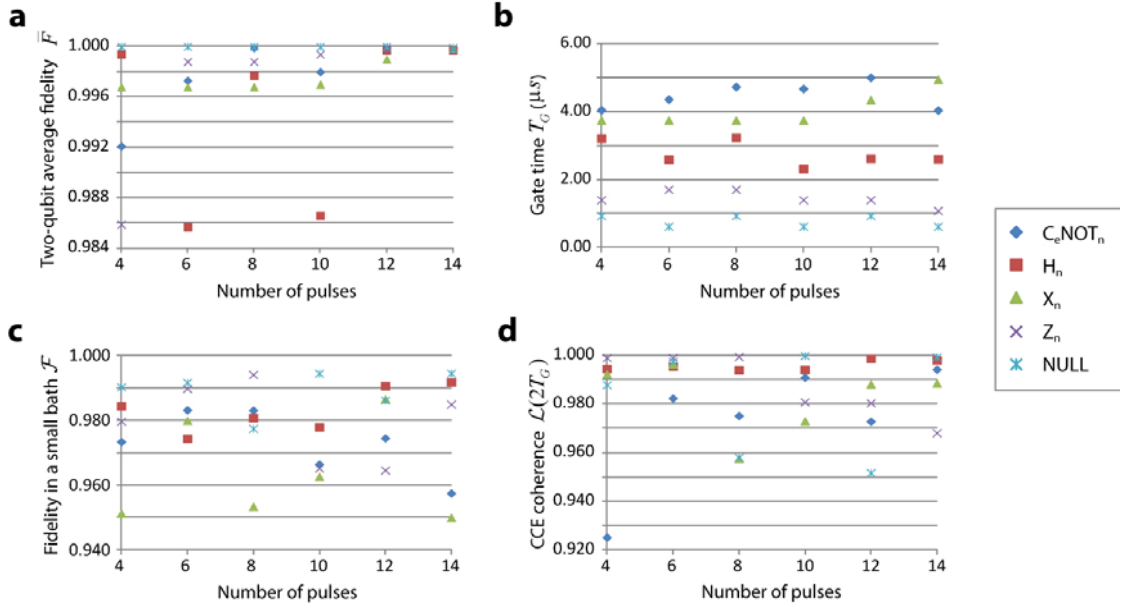

**Figure 2 | Simulation of two-qubit gates by dynamical decoupling.** By varying the timing parameters in the DD sequences with different number of pulses, 5 two-qubit gates are designed (Table S1). **a,** Ideal gate fidelities of the five gates from sequence optimization (no decoherence is included). The maximized fidelity approaches to unity when the number of pulses increases to 14. **b,** Operation times of the gates as designed in **a** as functions of the number of pulses. **c,** State fidelities evaluated by exact diagonalization of a small bath. While in general a high fidelity is achieved, the fidelities, in contrast to the optimized ideal fidelities in **a**, do not increase monotonically with the number of pulses. **d,** Coherence function probed after applying the gate twice as a function of the number of pulses.

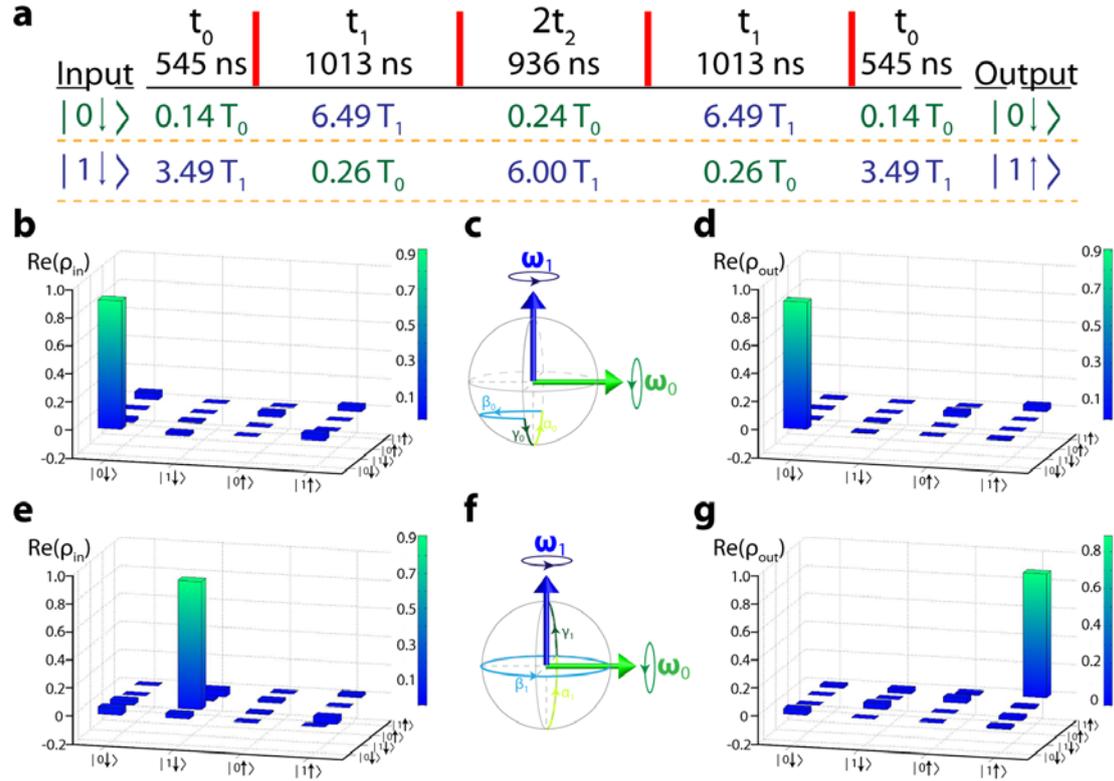

| Input | $t_0$ 545 ns | $t_1$ 1013 ns | $2t_2$ 936 ns | $t_1$ 1013 ns | $t_0$ 545 ns | Output |
|---|---|---|---|---|---|---|
| $\lvert 0\!\downarrow\rangle$ | $0.14\,T_0$ | $6.49\,T_1$ | $0.24\,T_0$ | $6.49\,T_1$ | $0.14\,T_0$ | $\lvert 0\!\downarrow\rangle$ |
| $\lvert 1\!\downarrow\rangle$ | $3.49\,T_1$ | $0.26\,T_0$ | $6.00\,T_1$ | $0.26\,T_0$ | $3.49\,T_1$ | $\lvert 1\!\uparrow\rangle$ |

**Figure 3 | Implementing the $C_eNOT_n$ gate by DD.** The effect of the gate was studied by performing state tomography on the system before and after the gate. **a,** The DD gate sequence demonstrated. **b-d,** Effect of the DD on the $\lvert 0\!\downarrow\rangle$ state. **b,** State tomography for the initial state $\lvert 0\!\downarrow\rangle$. Note that in all the state tomography results only the real parts are included. The imaginary parts are much smaller in amplitude and are included in the Supplementary Information. **c,** Schematic of evolution from $\lvert 0\!\downarrow\rangle$ steered by the DD . The nuclear spin first precesses about $\omega_0$ and alternates between $\omega_0$ and $\omega_1$ under the DD gate sequence. The nuclear qubit traces out the path $\alpha_0 - \beta_0 - (\gamma_0 + \alpha_0) - \beta_0 - \gamma_0$ under the sequence $t_0 - t_1 - (t_2 + t_2) - t_1 - t_0$ and returns to the original state at the end of the sequence. **d,** State tomography of the output state $U_{DD}\lvert 0\!\downarrow\rangle$. The resultant state fidelity was found to be 0.91. **e-g,** Similar to **b-d**, but for the initial state $\lvert 1\!\downarrow\rangle$. In the evolution steered

by the DD, the nuclear spin first precesses about $\omega_1$ for $t_0$. It then evolves during $t_1 - 2t_2 - t_1$ along the path $\alpha_1 - \beta_1 - \gamma_1$ and is hence flipped. The final state fidelity was measured to be 0.88.

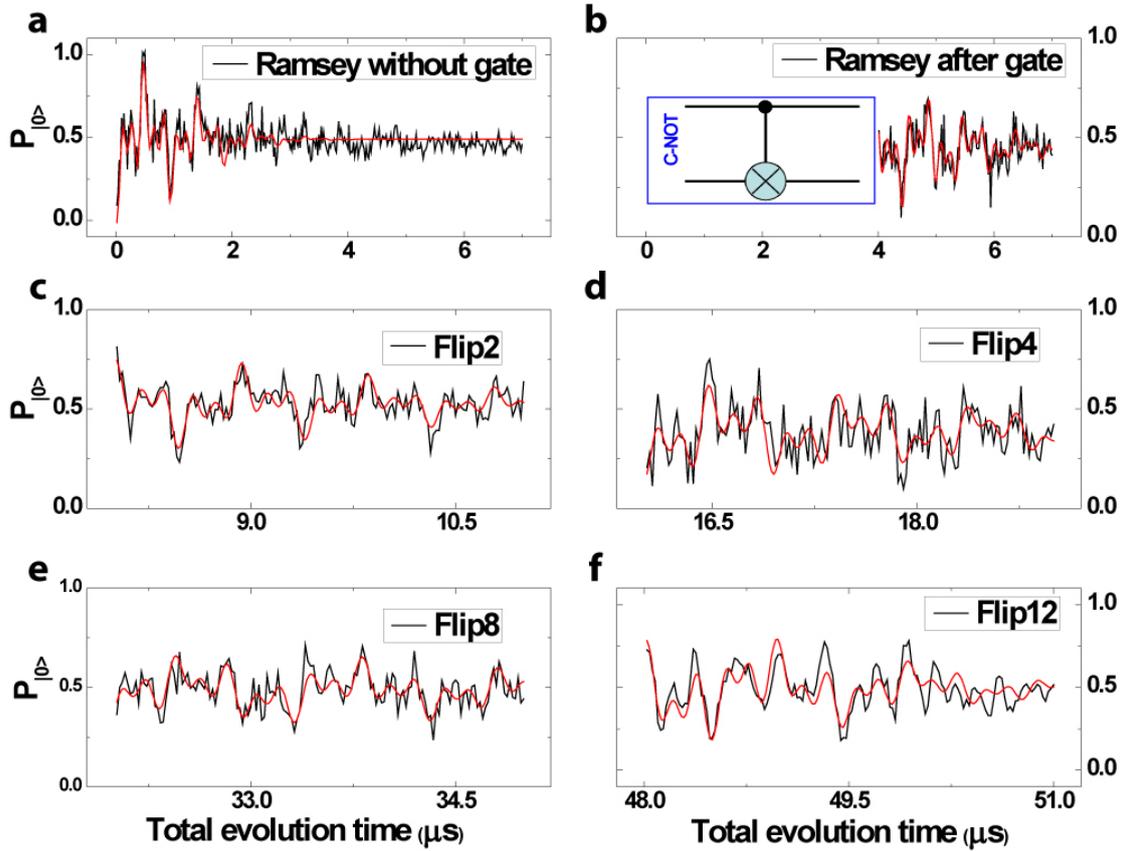

**Figure 4 | Coherence protection by the DD gate sequence.** The signals were measured by Ramsey interference before or after the DD control. **a ,** Centre spin coherence in FID without applying the DD pulses. The coherence was lost in $T_2^*$ =1.54(9) μs. **b,** Centre spin coherence after applying the DD gate sequence. Right after the DD gate sequence the centre spin coherence is 0, which evidences the maximum entanglement between the electron and the nuclear qubits by the $C_eNOT_n$ gate. The strong revival peaks in the Ramsey interference signal after implementing the DD gate indicates that the centre spin coherence was well protected. **c-f,** The centre spin coherence after the DD gate sequence was repeated for N=2, 4, 8, 12 times from c to f. Strong interference oscillations were observed in all cases, which indicates that the coherence time was elongated from $T_2^*$ by at least 30 folds.